\documentclass{PoS}

\title{Multiplicity, Transverse Momentum, Forward-Backward Long Range
  Corrrelations and Percolation of strings}

\ShortTitle{Correlations and string percolation}

\author{L.Cunqueiro, E.G.Ferreiro and \speaker{C.Pajares}\\
        Instituto Galego de F\'{\i}sica de Altas Enerx\'{\i}as and Departamento de F\'{\i}sica de Part\'{\i}culas, Universidade de Santiago de Compostela, 15782 Santiago de Compostela, Spain.\\
        E-mail: \email{pajares@fpaxp1.usc.es}}

\abstract{The behaviour of the normalized variance of the multiplictiy distribution and
the normalized transverse momentum fluctuations with  centrality is
naturally explained by the dependence on centrality of the number of
clusters of color sources. This dependence predicts a nonmonotonic behaviour
with centrality of the multiplicity associated to high $p_T$ events. The
clustering of color sources is also able to reproduce the increase with 
centrality of the forward-backward long-range correlations and its supression
compared to superposition model predictions.}

\FullConference{The 2nd edition of the International Workshop --- Correlations and Fluctuations in Relativistic Nuclear Collisions --- \\
                 July 7-9 2006\\
                 Galileo Galilei Institute, Florence, Italy}

\begin{document}

\section{Multiplicity and Transverse Momentum correlations}

 Event by event fluctuations of transverse momentum have been measured both at
 SPS and RHIC \cite{1,2,3,4,5,6,7}. The data show a nontrivial behaviour as a
 function of the centralitly of the collision. Concretely, the nonstatistical
 normalized fluctuations grow as the centrality increases, with a maximum
 around $Npart \simeq 100-150,$ followed by a decrease at larger
 centralities. The NA49 collaboration have presented their data on
 multiplicity fluctuations as a function of centrality for $Pb-Pb$ collisions
 \cite{8,9}. A nonmonotonic centrality (system size) dependence of the
 multiplicty scaled variance was found. Its behaviour is similar to the
 one obtained for the $\Phi (p_{T})$ measure, used by the 
$NA49$ Collaboration  to quantify the $p_{T}$ fluctuations \cite{2}, suggesting
 that they are related to each other \cite{11}. The $\Phi$ measure is
 independent of the distribution of the number of particle sources if the
 sources are identical and independent of each other. This implies that $\Phi$
 would be independent of the impact parameter if the nucleus-nucleus collision
 was a simple superposition of nucleon-nucleon interactions.

In the framework of string clustering \cite{12} such a  behaviour is naturally
explained \cite{13,14}. Let us remember the main features of the model. In a
nucleus-nucleus collision, color strings are stretched between partons from
the projectile and the target. This strings decay into new strings by $q-\overline{q}$
pair production and finally hadronize to produce the observed particles. For
the decay of the strings we apply the Schwinger mechanism of fragmentation,
where the decay is controlled by the string tension that depends on the color
field of the string. 

The strings have longitudinal and transverse dimensions, and the density of
created strings in the first step of the collision depends on the energy and
on the centrality of the collision. One can consider the number of strings
$N_{S}$ in the central rapidity region as proportional to the number of
collisions, $N_{A}^{\frac{4}{3}}$, whereas in the forward and backward region it
becomes proportional to the number of participants $N_{A}$. (We follow the dual
parton model \cite{15,16} or the quark gluon string model \cite{17}). We will use the variable 
\begin{eqnarray}
\eta = N_{S} \frac{S_{1}}{S_{A}} \label{eq:eta} 
\end{eqnarray}  
proportional to the density of strings, where $S_{A}$ corresponds to the
nuclear overlap area, $S_{A}=\pi R_{A}^2$ for central collisions, and $S_{1}$,
to the area of one string, $S_{1}=\pi r_{0}^2$ ($r_{0} \simeq 0.2-0.3$
fm). With the increase of energy and/or atomic number of the colliding nuclei,
the density grows, so the strings begin to overlap forming clusters. We assume
that a cluster of n strings that occupies an area $S_{n}$ behaves as a single
color source with a higher color field, generated by a higher color charge
$\vec{Q_{n}}$. This charge corresponds to the vectorial sum of the color
charge of each individual string $\vec{Q_{1}}$ The resulting color field
covers the area $S_{n}$ of the cluster. As $Q_{n}^2=(\sum_{i=1}^{n}\vec{Q_{1}})^2$ and since the individual string colors may be
arbitrarily oriented, the average $\vec{Q_{1i}}\vec{Q_{1j}}$ is zero and
therefore, $Q_{n}^2=nQ_{1}^2$ if the strings fully overlap. Because the
strings may overlap only partially we introduce a dependence on the area of the cluster, 
\begin{eqnarray}
 Q_{n}=\sqrt{\frac{nS_{n}}{S_{1}}}Q_{1}. \label{eq:algeb} 
\end{eqnarray}

Note that if the strings are just touching each other, $S_{n}=nS_{1}$ and
$Q_{n}=n Q_{1}$, so the strings behave independently. On the contrary, if
they fully overlap, $S_{n}=S_{1}$ and $Q_{n}=\sqrt{n}Q_{1}$. Knowing $Q_{n}$,
one can compute the multiplicity $\mu_{n}$ and the mean transverse momentum
$<p_{T}>_{n}$ of the particles produced by a cluster of n strings
\cite{17,18}. According to the Schwinger mechanism for the fragmentation of
the clusters, one finds
\begin{eqnarray} <\mu>_{n}=\sqrt{\frac{nS_{n}}{S_{1}}}<\mu>_1 ,\; 
\; \: <p_{T}>_{n}=(\frac{nS_1}{S_{n}})^{\frac{1}{4}} <p_T>_1
\label{eq:Schwinger} 
\end{eqnarray}
where $<\mu>_{1}$ and $<p_{T}>_1$ correspond to the mean multiplicity and the mean
 transverse momentum of the particles produced by one individual string. As the
 energy and/or the number of paricipants of the collision increase, the
 density of strings increases. At a certain critical density ($\eta_{C} \simeq
 1.2-1.5$, depending on the nuclei-profile used) a macroscopical cluster
 appears wich marks the percolation phase transition, which is a second order,
 nonthermal phase transition. (The formation of a macroscopial cluster of
 strings can be seen as due to multiple partonic interactions, which can
 approximately give rise to a thermal spectrum. In this way, the critical
 percolation density is related to a critical temperature \cite{18}.)

To obtain the mean $p_{T}$ and the mean multiplicity of the collision at a
given centrality one needs to sum over all formed clusters and to average
over all events
\begin{eqnarray}
<\mu>=\frac{\sum_{i=1}^{N_{events}} \sum_{j} <\mu>_{nj}}{N_{events}}, \;
\;<p_T>=\frac{\sum_{i=1}^{N_{events}}\sum_{j}
  <\mu>_{nj}<p_T>_{nj}}{\sum_{i=1}^{N_{events}}\sum_{j}<\mu>_{nj}}.
\label{eq:distrib} 
\end{eqnarray}

The sum over j goes over all individual clusters j, each one formed by $n_{j}$
strings and occupying an area $S_{nj}$. The quantities $n_{j}$ and $S_{nj}$
are obtained for each event, using a Monte-Carlo code \cite{16}, based on the
quark gluon string model. With our code, once we fix the energy and the nuclei
of the collision, we obtain, for each event, a number of participant
nucleons and a  configuration for the created strings. Each string is
generated at an identified impact parameter in the transverse space. Knowing
the transverse area of each string, we identify all the clusters formed in
each event, the number of strings $n_j$ that conforms a cluster $j$ and the
area occupied by each cluster. We use a Monte-Carlo code for the cluster
formation to compute the number of strings that come into each cluster and
the area of the cluster. Conversely, we do not use a Monte-Carlo code for the
decay of the cluster because we apply analytical expressions
(eq. (\ref{eq:distrib})). We assume that the multiplicity distribution
of each cluster follows a Poissonian of mean value $<\mu>_{nj}$ and therefore
the variance $<\mu^2>_{nj}-<\mu>_{nj}^2$  is $<\mu>_{nj}$.
It is easy to see that at low densities the scaled variance is given by

$$\frac{Var(\mu)}{<\mu>}=1+<\mu>_1$$

and at high densities, $$\frac{Var(\mu)}{<\mu>} \longrightarrow 1.$$

\begin{figure}
\includegraphics[width=.6\textwidth]{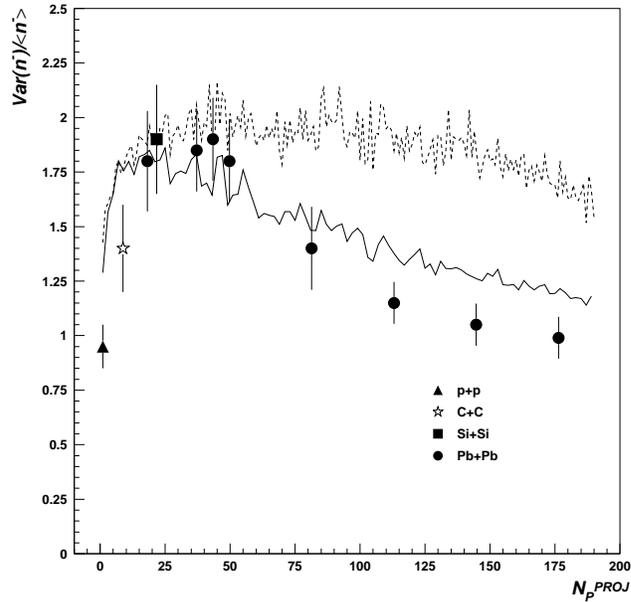}
\caption{Our results for the scaled variance of negatively charged particles
  in Pb-Pb collisions at SPS energies compared to NA49 data. Solid
  line: clustering of colour sources. Dashed line: independent strings.}

\label{fig1}
\end{figure}

Our results for the scaled variance for negative particles are presented in
fig.(\ref{fig1}). The rapidity interval is  $4.0<y<5.5$. We have also included our
results without cluster formation. One can observe that when clustering is
included we find a good agreement with the experimental data. We see that the
clustering produces a decrease of the scaled variance for central collisions,
where the density of strings increases and the clustering has a bigger
effect. At RHIC energies our results are similar to the ones obtained at SPS
energies. The fluctuations on the number of target participants at a fixed
number of proyectile participants have been pointed out to be an important
contribution to the scaled multiplicity variance  \cite{10,19}. There are many
string models which find no agreement with  data because such fluctuations do
not contribute to the projectile rapidity hemisphere (HIJING,URQMD,HSD). In
these models, there is momentum exchange but no color exchange between partons
of the projectile and the target. In DPM or QGSM there is color exchange and
the strings connect both rapidity hemispheres. Although at very high density
the scaled variance goes to one, our result is above one and also
clearly above the experimental data. (The new NA49 data is clearly below one
\cite{10}). The reason for this difference is that we do not take into account
the energy conservation in the formation of clusters due to the use of
analytical formulae. In fact, in the forward rapidity range considered, at a
large fixed number of projectile participants, the energy conservation implies
that the number of strings and also the number of target participants are
almost fixed in such a way that the scaled variance at high centrality is more
supressed. This effect should be weaker at mid rapidity.

The PHENIX Collaboration \cite{20} has measured the centrality dependence of
the transverse momentum fluctuations using the observable $F_{P_{T}}$ that
quantifies the deviation of the observed fluctations from statistically
independent particle emission: 
\begin{eqnarray}
  F_{P_{T}}=\frac{\omega_{data}-\omega_{random}}{\omega_{random}}
  \label{eq:fpt} 
\end{eqnarray}
where 
\begin{eqnarray} \omega =\frac{\sqrt{<p_T>^2-<p_T>^2}}{<p_T>}. \label{eq:omega} 
\end{eqnarray}

\begin{figure}
\includegraphics[width=.6\textwidth]{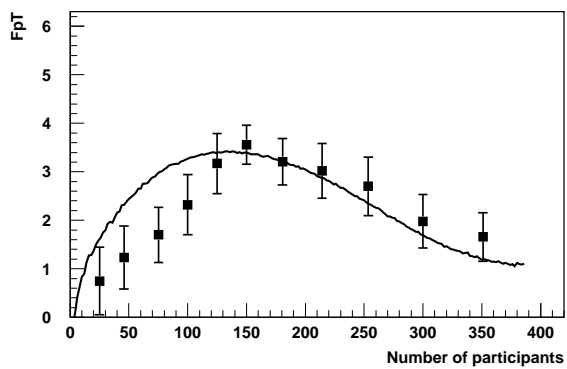}
\caption{$F_{P_{T}}$ versus the number of participants. Experimental data
from PHENIX at $\sqrt{s}=200 GeV$ are compared to our results(solid line).}
\label{fig2}
\end{figure}

The comparison of our results for the dependence of $F_{P_{T}}$ on the number
of participants $N_p$ to the PHENIX data is shown in fig.(\ref{fig2}). An acceptable
agreement is obtained.

The behaviour of the transverse momentum fluctuations as well as the behaviour
of the multiplicity fluctuations can be understood as follows: at low density, most of the
particles are produced by individual strings with the same $<p_T>$ and $<\mu>$,
so the fluctuations are small. Similarly, at large density above the
percolation critial point there is essentially only one cluster formed by most
of the strings created in the collision and therefore fluctuations are not
expected either. 

\begin{figure}
\includegraphics[width=.6\textwidth]{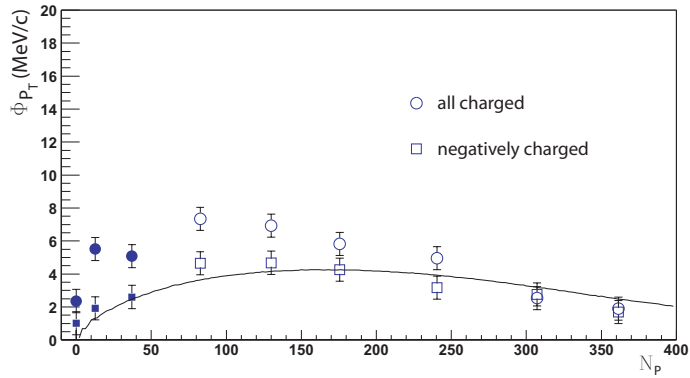}
\caption{ $\Phi_{P_{T}}$ versus the number of participants. Experimental data
from NA49 Collaboration at SPS energies compared to our results(solid line).}

\label{fig3}
\end{figure}

In fig.(\ref{fig3}) our results for the observable $\Phi_{P_{T}}$ of
charged particles in Pb-Pb central collisions at 158AGeV are compared to the
data of NA49 \cite{2}. A good agreement is obtained.

PHENIX and STAR Collaborations have pointed out (\cite{21,22}) minijets as the
main source of transverse momentum fluctuations. In our approach, these
fluctuations have the same origin as the multiplicity fluctuations, namely the
clustering of color sources. Since a cluster of strings produces particles
with a harder $p_T$ spectrum than in the unclustering case, our approach is
compatible with the role of minijets in transverse momentum fluctuations at
RHIC energies. Notice that at SPS energies there are transverse momentum
fluctuations although the production of minijets is negligible. More studies
on both $p_T$ and $\Phi$ spaces would be very convenient, as it has been
emphasized at this workshop \cite{22}.

Finally let us mention that our results are consistent with the clustering
analysis\cite{23} presented at this workshop. 

\vspace{1 cm}

\section{Multiplicity associated to high $p_{T}$ events and multiplicity fluctuations}

The events which are self-shadowed have singular properties concerning the
multiplicity distribution associated to them. We call self-shadowed events
in hadron-hadron, hadron-nucleus or nucleus-nucleus collisions those events
whose inelastic cross section depends only on the elementary cross section for
such events \cite{24}. Assuming that hadron-hadron, hadron-nucleus and
nucleus-nucleus collisions are a superposition of independent elementary
collisions, it is shown in \cite{25} that the multiplicity distribution
associated to self-shadowed events, $P_{C}(n)$ is approximately given in terms
of the total multiplicity distribution $P(n)$ by 
\begin{eqnarray}
 P_{C}(n) \simeq \frac{nP(n)}{<n>}. \label{eq:universal}
\end{eqnarray}

There are many different self-shadowed events, for instance non-diffractive
events, anihilation events in $\overline{p}A$ collisions, nonisolated fast
baryons in pA or AA collisions, or high $p_{T}$ events in hh, hA or AA
collisions. Equation (\ref{eq:universal}) has been checked in high energy
pp collisions for the multiplicity distribution associated to $W^{+ -}$ and 
 $Z^0$ production and also for the multiplicity distribution associated to jet
production and annihilation \cite{25}. In nucleus-nucleus collisions, data of
ISR experiments \cite{26} on events with $p_{T} \ge 3GeV$ produced in
    $\alpha-\alpha$ collisions also satisfy eq.(\ref{eq:universal}).

The eq.(\ref{eq:universal}) has been obtained assuming independent
superposition of elementary interactions. This assumption is not justified
for heavy nuclei collisions at RHIC energies where collective
interactions, such as
percolation of strings, are at work. However, it can be argued \cite{27} that
even in these cases, eq. (\ref{eq:universal}) is approximately valid,
considering the collision as a superposition of different clusters of
elementary interactions (strings). Since high $p_{T}$ events are self-shadowed,
from eq. (\ref{eq:universal}), we can write the difference between the average
multiplicity associated to high $p_{T}$ events $<n>_{C}$ and the total
average multiplicity in terms of the scaled variance of the total multiplictiy
distribution \cite{27}:  
\begin{eqnarray}
 <n>_C-<n>=\frac{<n^2>-<n>^2}{<n>}.    \label{eq:relation} 
\end{eqnarray}

The equation (\ref{eq:universal}) can be easily checked experimetally.
\vspace{2cm}

\section{Forward-Backward Long Range Correlations}

In any model based on a superposition of elementary and statistically
independent collisions, the squared forward-backward dispersion is proportional to the
square dispersion of the number of elementary collisions \cite{28}. In fact,
we have

\begin{eqnarray} D^2_{FB} \equiv <n_F
n_B>-<n_F><n_B>=<N>(<n_{0F}n_{0B}>-<n_{0F}><n_{0B}>)+ \nonumber \\
(<N^2>-<N>^2) <n_{0F}><n_{0B}>  \label{eq:correlations} \end{eqnarray}

where N stands for the number of elementary interactions, $n_{0F}(n_{0B})$ for
the number of forward(backward) produced particles in an elementary
interaction and $n_{F}(n_{B})$ for the total number of forward(backward)
particles.

The first term of (\ref{eq:correlations}) is the correlation between particles
produced in the same elementary interaction. Assuming these  correlations to
have short range in rapidity, this term vanishes if one takes a rapidty gap
larger than $1-1.5$ units between the forward and backward rapidity
intervals. In this way, one is left with the last term in
(\ref{eq:correlations}). We see that there is a long range correlation between
particles which are far away in rapidity. This correlation is due to the
fluctation in the number of elementary interactions, controlled by
unitarity. This term increases with the number of elementary interactions,
therefore we expect that the long-range correlations increase with  energy
and the size of the nucleus in hh, hA and AA collisions. However, if there are
interactions among strings, the number of independent elementary interactions
translates approximately into the number of clusters of strings. Therefore a
clear supression of long range correlations relative to the expected in  a
superposition picture is predicted \cite{29,30}.

\begin{figure}
\includegraphics[width=.6\textwidth]{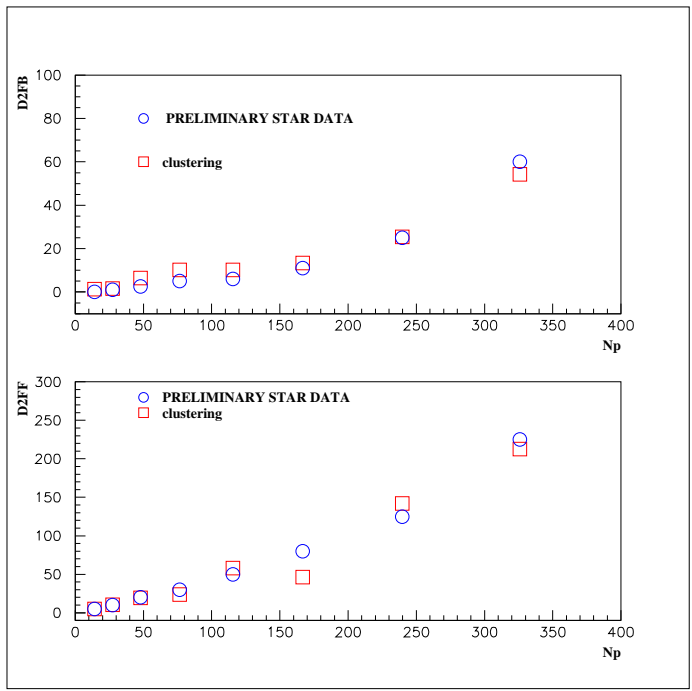}
\caption{$DFB^2$ and $DFF^2$ versus the number of participants compared to PRELIMINARY STAR data}

\label{fig4}
\end{figure}

The preliminary data of STAR presented in this workshop \cite{31} show that
in fact there is a strong supression of long range correlations. In fig.(\ref{fig4}) we
compare the preliminary data, obtained with a rapidity gap of $1.6$ units in
the central rapidity region, and a forward and backward intervals of $0.2$ units,
to our results \cite{32} of percolation of strings. A good agreements is
obtained.

\begin{figure}
\includegraphics[width=.6\textwidth]{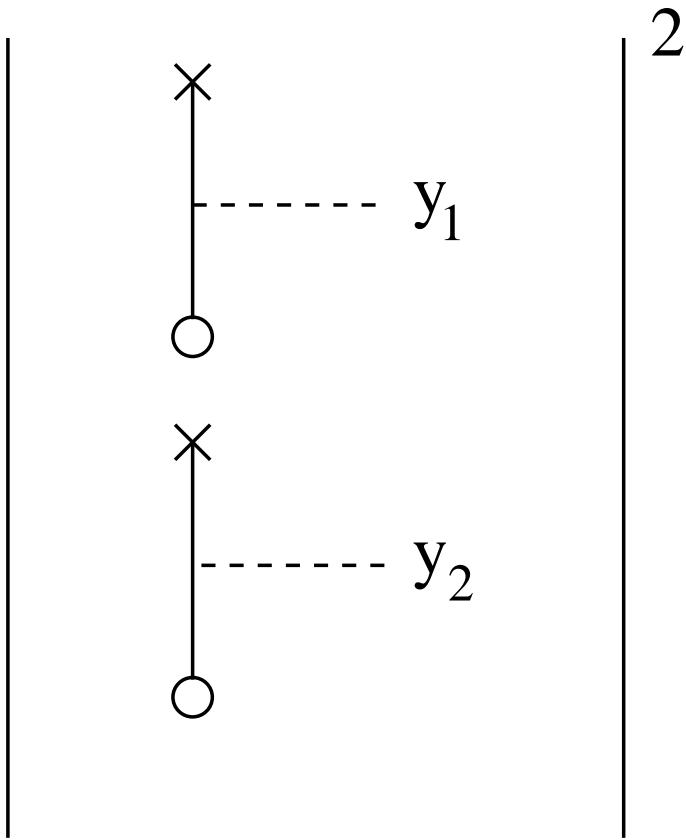}
\caption{ The leading order diagram which induce long range correlation in
  rapidity. The source of one nucleus is given by the x and that of the other
 by the O. The produced gluon is denoted by the dotted line.}

\label{fig5}
\end{figure}

Finally, let us mention that the Color Glass Condensate (CGC) generates
distintive predictions for the long-range component of the correlations
\cite{33}. The main contribution to this component is given by the diagram of
fig.(\ref{fig5}) which is 

\begin{eqnarray}
 <\frac{dN_{F}}{dy_1}\frac{dN_{B}}{dy_2}>=<(\frac{dN}{dy})^2> \sim
\frac{\pi R^2 Q_S^2}{\alpha_S^2} \sim \frac{1}{\alpha_S}\frac{dN}{dy}
\label{eq:CCG} 
\end{eqnarray}

where $Q_S$ is the saturation momentum.

\begin{figure}
\includegraphics[width=.6\textwidth]{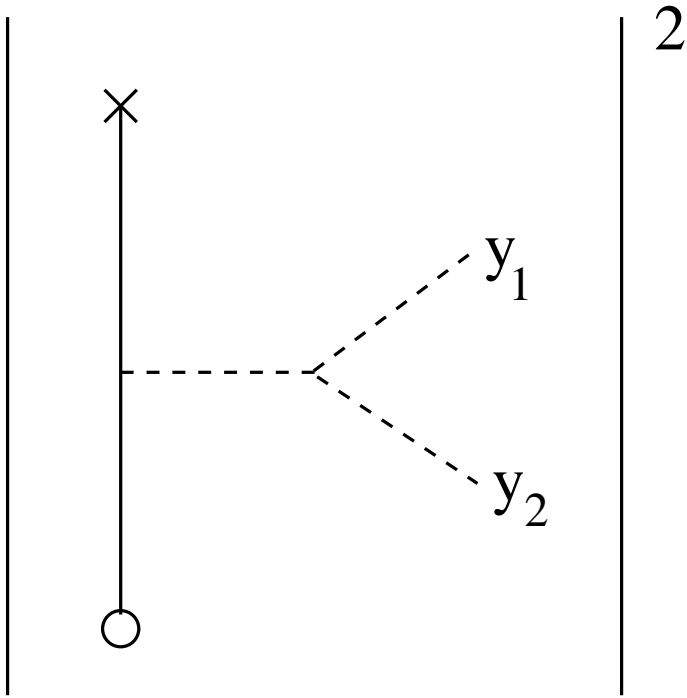}
\caption{The leading order diagram which induce short range correlation in rapidity. Here two gluon from the different sources scatter and produce two
gluons in the forward and backward hemisphere.}

\label{fig6}
\end{figure}

On the contrary, the main contribution to the short range correlation is given
by diagram of fig.(\ref{fig6}). This diagram has two factors of $\alpha_S$ and should
give a contribution to the total multiplicity fluctuations of order
\begin{eqnarray}
 <\frac{dN}{dy_1}\frac{dN}{dy2}> \sim \alpha_S \pi R^2 Q_S^4. \label{eq:order}
\end{eqnarray}
The different powers of $\alpha_S$ in eqs. (\ref{eq:CCG}) and (\ref{eq:order})
allow us to easily disentangle the long-range correlations from the short range
correlations. The predictions of CGC are not very different from the
percolation of strings ones, what is not unexpected given the similarities
between both approaches.

We thank the organizers for such a nice meeting. This work was done under
contract FPA2005-01963 of CICYT of SPAIN.

\end{document}